\documentclass[aps,prb,twocolumn,superscriptaddress]{revtex4}

\usepackage{mathrsfs}
\usepackage{graphicx}
\usepackage{color}
\usepackage{amsmath}
\usepackage{amssymb}

\begin{document}

\title{Kinetics of polymer looping with macromolecular crowding: \\ effects of volume fraction
and crowder size}

\author{Jaeoh Shin}
\affiliation{Institute for Physics \& Astronomy, University of Potsdam, D-14476
Potsdam-Golm, Germany}

\author{Andrey G. Cherstvy}
\affiliation{Institute for Physics \& Astronomy, University of Potsdam, D-14476
Potsdam-Golm, Germany}

\author{Ralf Metzler}
\email{rmetzler@uni-potsdam.de}
\affiliation{Institute for Physics \& Astronomy, University of Potsdam, D-14476
Potsdam-Golm, Germany}
\affiliation{Department of Physics, Tampere University of Technology, FI-33101
Tampere, Finland}


\begin{abstract}
The looping of polymers such as DNA is a fundamental process in the molecular
biology of living cells, whose interior is characterised by a high degree of
molecular crowding. We here investigate in detail the looping dynamics of
flexible polymer chains in the presence of different degrees of crowding. From
the analysis of the looping-unlooping rates and the looping probabilities of the
chain ends we show that the presence of small crowders typically slow down the chain
dynamics but larger crowders may in fact facilitate the looping. We rationalise
these non-trivial and often counterintuitive effects of the crowder size onto
the looping kinetics in terms of an effective solution viscosity and standard excluded
volume effects. Thus for small crowders the effect of an increased viscosity
dominates, while for big crowders we argue that confinement effects (caging)
prevail. The tradeoff between both trends can thus result in the impediment or
facilitation of polymer looping, depending on the crowder size. We also examine
how the crowding volume fraction, chain length, and the attraction strength of the
contact groups of the polymer chain affect the looping kinetics and hairpin
formation dynamics. Our results are relevant for DNA looping in the absence and
presence of protein mediation, DNA hairpin formation, RNA folding, and the folding
of polypeptide chains under biologically relevant high-crowding conditions.
\end{abstract}

\maketitle

Abbreviations: MMC, macromolecular crowding; PDF, probability density function; LJ, Lennard-Jones; FENE, finitely-extensible non-linear elastic; PEG, polyethelene glycol; ssDNA, single-stranded DNA; dsDNA, double-stranded DNA; MW, molecular weight; MSD, mean squared displacement.

\section{Introduction}

Molecular reactions in living biological cells are running off in a
highly complex environment, that is compartmentalised by membrane structures and
crowded with macromolecules and structural cytoskeletal networks. Macromolecular
crowding (MMC) makes up a ``superdense'' \cite{golding} environment modulating
the kinetics of various biochemical processes in cells. Inter alia, this mechanism
is employed biologically to tune the DNA accessibility in the cyto- and nucleoplasm.
MMC non trivially influences the levels of gene expression, and the size of the
crowders dramatically modifies the response of genetic elements \cite{leduc13}.
In particular, it was found that in solutions of small crowders the rate of gene
expression only varies slightly with the volume fraction $\phi$ of the crowders,
while large crowders boost the expression levels many-fold \cite{leduc13}. 

More
specifically, MMC constitutes a non-specific environment controlling the looping
properties of biopolymers such as nucleic acids and polypeptides. Polymer looping
is indeed a ubiquitous mechanism of DNA protection, compaction, and gene regulation
in both bacteria and higher organisms \cite{philips}. DNA looping is vital for the
regulation of transcription and effects the robustness of bio-switches
\cite{allen09}. The effects of MMC on kinetics of DNA looping are of paramount
importance for the speed, efficiency, and precision of gene regulatory networks
\cite{leduc13,elf09,szleifer14}. Inspired by the impressive body of experimental
evidence for the relevance of MMC on biochemical processes, we here scrutinise
the key role of the crowder size for the kinetics and thermodynamics of polymer
looping.

The quantitative study of the diffusion-limited encounter of the end monomers
of a polymer chain in a mixture of crowders of varying sizes and the analysis
of the effective viscosity of the solution is a formidable theoretical problem.
Despite the progress of the understanding of polymer looping and cyclisation at
dilute solvent conditions by theoretical approaches \cite{fixman74,loop-theory-1,
loop-theory-1a,
loop-theory-2} and by simulations \cite{davide06crowd,shin12looping,loop-simul-1,
loop-simul-2,loop-simul-3,langowski-looping-zimm-reactive-groups,heermann-sm-looping}, polymer looping
in the presence of MMC \cite{loop-crowd-simul-1,loop-crowd-simul-3,
loop-crowd-simul-3a} still poses a number of challenges, which are our main
targets here.

\begin{figure}
\begin{center}
\includegraphics[width=5.5cm]{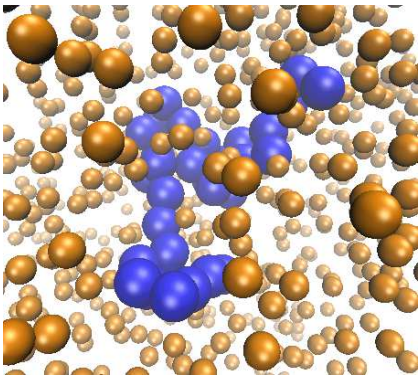}
\end{center}
\caption{Typical polymer conformation in the presence of MMC. The polymer chain
(blue spheres) consists of $n=32$ monomers, the fraction of crowders (golden
spheres, rendered smaller for better visibility of the polymer) is $\phi=0.1$,
and the size of the crowders is $d_{\text{cr}}=1\sigma$ in terms of the monomer diameter $\sigma$ of the polymer chain. Video-files illustrating the dynamics looping
dynamics of polymer chains for small and big crowders are included in the
Supporting Information.
\label{fig1}}
\end{figure}

\begin{figure}
\begin{center}
\includegraphics[width=7cm]{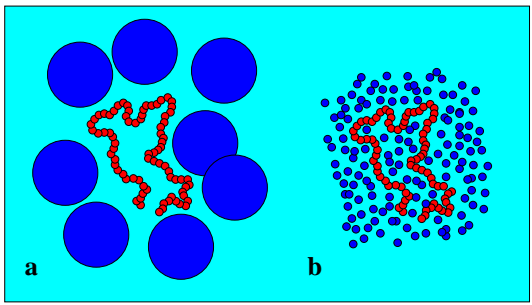}
\end{center}
\caption{Crowder size effect: Large crowders lead to the caging of the polymer (a), while small crowders tend to mix with the chain monomers (b) and increase the effective viscosity. Note that the size of the chain is the same in both images.
\label{fig6}}
\end{figure}

It is known that MMC generally facilitates the association of proteins via
volume exclusion effects and favours more compact states \cite{minton08}. Polymer
looping, however, involves the diffusion of an extended and chain length-dependent
fragment of the polymer in crowded solutions. This non-locality effect renders the
trends of the inhibition or facilitation of polymer looping kinetics in the
presence of MMC less intuitive. Looping is a fundamental dynamic property of
polymers which can be directly probed by methods such as fluorescence energy
transfer \cite{brauchle}. A comprehensive theory of polymer looping under crowded
conditions is not straightforward. We here employ extensive crowder-explicit
simulations of polymer looping including a number of important physical and biochemical
ingredients.

Polymer organisation in the presence of MMC and spatial confinement is a common
theme in biophysics \cite{italian-mafia-11review}. It affects, for instance, the
segregation of DNA rings in dividing bacteria cells \cite{stasiak-chrom,shin14rings} as well as
the territorial organisation of
DNA inside eukaryotic nuclei \cite{cremer-chrom-terr} and bacteria
\cite{mirny13bacterial-chmorosomes}. Of particular interest is polymer looping and
knotting in MMC-dominated solvents \cite{denton14,franosch13,weiss14}. The highly
crowded environments of real biological cells feature volume occupancies of up
to $\phi\sim $30\% \cite{crowd2,elcock-ecoli-cytoplasm}. In vitro,
concentrated solutions of naturally occurring proteins, globular and branched polymers (lysozyme, serum albumin, PEG, dextran, Ficoll,
etc.) mimic MMC conditions in a more controlled environment
\cite{szymanski,pan}. On top of MMC volume exclusion, the eukaryotic cytoskeleton
forms a spatial mesh with a period of several tens of nm affecting the
diffusion of cellular components.

Excluded-volume interactions by crowders favour molecular association reactions
\cite{vietnam13}, speed up the folding of proteins into their native structures
\cite{schreiber,dortmund14,thirum05,thirum11nature,folding-pnas}, and facilitate
the assembly of virus capsids \cite{mateu-hiv}. The effects of the crowder size
were studied for polypeptide folding \cite{denesyuk11} and protein fibrillisation
\cite{vietnam13}. We note that apart from MMC in the cytosol of biological cells,
crowding is also an important ingredient for the diffusional dynamics of embedded
proteins and lipid molecules in biological membranes \cite{weigel,jae,jae1}. Also note that the thermodynamics and the demixing transitions in the
mixtures of colloidal particles and linear polymers have been explored
\cite{france-mixture}, in particular in the limit of long polymers (the so-called
"protein limit") \cite{greek-polymers}.

The biological relevance for the study of polymer looping is due to its central
role in gene regulation, for instance, in the formation of DNA loops induced by
transcription factor proteins such as Lac or $\lambda$ repressor \cite{loopsDNAlac,
ptashne2014loosLAC,hensel}. Inter-segmental protein jumps along DNA made possible
via looping facilitate protein diffusion in DNA coils \cite{metzler08pnas,
metzler09pnas} and affects MMC-mediated gene regulation \cite{leduc13,elf09,wolde14,
holyst14}. Another example is the dynamics of the DNA chain itself on
various levels of DNA structural organisation ranging from the bare DNA, via
chromatin fibres, to complex chromosomal filaments \cite{philips,teif13}. We also mention
protein- \cite{protein-folding-crowd-confin-review} and RNA-folding \cite{dupuis}
reactions. 

Experimentally, the effects of polymeric crowders onto the
opening-closing dynamics of ssDNA hairpins with complementary sticky ends
\cite{pnas-hairpins-kruchevsky-well-cited,anom-dyn-DNA-hairpins-PRL} were
studied in detail \cite{weiss13crowd}. It was demonstrated in
Ref.~\cite{weiss13crowd} that
ssDNA hairpin formation dynamics is dramatically slowed down in highly-crowded
solutions of dextran and PEG of varying molecular weights (MWs), MW$\sim$ 0.2-10 kDa.
Also, the fraction of open hairpins gets reduced substantially by relatively
large crowders, in contrast to low-MW solutions of sucrose. In the latter,
the similarly slowed-down DNA hairpin dynamics due to a higher viscosity
of the medium, the fraction of hairpins stayed nearly constant with
crowding. Note that the experimental setup of Ref. \cite{weiss13crowd}
only allowed to measure the geometric average of looping-unlooping times
$\tau_K$. A separate measurement of looping  $T_{l}$ and unlooping  $T_{ul}$
times of the cohesive chain ends as a function of MMC fraction $\phi$ was
not feasible. The fraction of time the hairpins are in a looped state was also measured \cite{weiss13crowd}.

Some effects of MMC on polymer looping were analysed recently \cite{davide06crowd,
loop-crowd-simul-3}. For instance, for \textit{implicit\/} attractive depletion potentials
between polymer segments (mimicking MMC) the polymer looping ($T_l$) and unlooping
($T_{ul}$) times (see below) for  $\phi=0.15$ and fixed size of crowders were quantified by
simulations \cite{davide06crowd}. For long chains, the increase of the looping
time $T_{l}$ obeys the scaling relation
\begin{equation}
\label{eq-tl}
T_l(n)\sim n^{2\nu+1}\sim n^{2.2}
\end{equation}
with the chain length $l=n\sigma$ \cite{davide06crowd}. Here $\nu\approx3/5$ is the
Flory
exponent \cite{grosberg}. Relation \eqref{eq-tl} is indeed supported by polymer
cyclisation theory \cite{loop-simul-3}. Experimentally, the rate of formation of
DNA hairpins drops somewhat faster with the chain length, $T_l(n)\sim n^{2.6\pm
0.3}$, probably due to excluded-volume effects
\cite{pnas-hairpins-kruchevsky-well-cited}. Moreover, it was predicted that due to a non-trivial interplay of the enhanced solution viscosity and polymer "crumpling" the looping time varies non-monotonically with $\phi$ \cite{davide06crowd}.

\begin{figure}
\includegraphics[width=9cm]{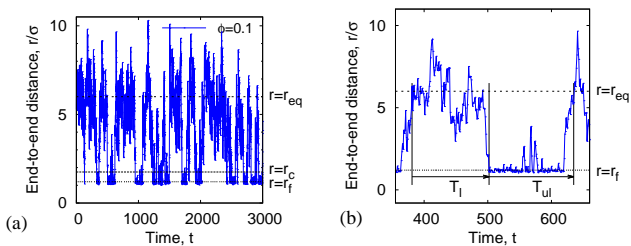}
\caption{Time evolution of the polymer end-to-end distance and definition of the
 looping-unlooping $T_{l,ul}$ and opening-closing $T_{op,cl}$. The equilibrium
$r_{\text{eq}}$ and critical $r_c$ distances are indicated. The simulated chain
consists of $n=16$ monomers, and the crowding fraction is $\phi=0.1$.
\label{fig2}}
\end{figure}

In contrast, the unlooping time $T_{ul}$ exhibits only a weak dependence on
the chain length \cite{davide06crowd}. A finite cohesive energy of polymer
ends, $\epsilon_s>0$, gives rise to more extended ``looped'' periods and longer
unlooping times \cite{davide06crowd}. The looping time, the time separating
the extended and looped states of the chain, becomes shorter due to
``depletion-based crowding'' for longer chains, i.e., more compact polymer states
are favoured, effecting a slow-down of the unlooping dynamics \cite{davide06crowd}.

We here report results from extensive Langevin dynamics simulations of the
looping of Rouse-like flexible polymers in solutions of \textit{explicit} nearly
hard-sphere crowders (see Fig.~\ref{fig1}). We examine the effects of the
crowding volume fraction
$\phi$, the crowder diameter $d_{cr}$, the stickiness $\epsilon_s$ of the
end monomers, and the chain length $n\sigma$, where $\sigma$ is the monomer
diameter. We showed recently \cite{shin14rings}
that for two polymer \textit{rings} under confinement and crowding conditions the contact
properties are non-monotonic in the crowding fraction $\phi$. Here, we demonstrate
that MMC has unexpected effects on the looping dynamics as well, due to competition between depletion effects facilitating looping and
an increased effective solution viscosity slowing down the looping kinetics, see
Figs. \ref{fig1}, \ref{fig6}.

\section{Model and Methods}
\label{sec-model}

To study polymer-nanoparticle mixtures by computer simulations,
Monte-Carlo and Molecular Dynamics
investigations were
conducted in the literature to elucidate the static and dynamical behaviour of binary mixtures
of polymers and crowders. Important ingredients were included
in simulations to render the results applicable to realistic situations, for
instance, in cells. Thus, the effects of compressible polymers \cite{denton11},
non-spherical \cite{kudlay12} and charged crowding nanoparticles
\cite{sdenton14,denton05} onto polymer-crowder demixing as well as
the implications of confinement \cite{looping-confinement,johner06}
and viscoelastic effects \cite{chera12visco} on polymer looping kinetics
were studied.

Computer simulations \cite{davide06crowd} revealed e.g. that the unlooping time
$T_{ul}$  stays nearly constant with $n$ and increases 3-4 times as crowding
fraction grows from $\phi=$0 to 0.15. Note that because
of a limited applicability of the effective depletion potentials used, only
moderate $\phi$ values were studied in Ref.~\cite{davide06crowd}. The unlooping
time $T_{ul}$ is defined in our study as the time required for the chain to
expand from the close-end to the equilibrium state, somewhat different from
the definition used in Ref. \cite{davide06crowd}, see Fig. \ref{fig2}.

\subsection{Potentials and Approximations}

Performing Langevin dynamics simulations of flexible polymers, we here examine
the looping probabilities of the chain ends in the presence of MMC. The polymer
chain is modelled within bead-spring model with finitely extensible nonlinear
elastic (FENE) potentials,
\begin{equation}
U_{\text{FENE}}(r)=-\frac{k}{2}r_{\text{max}}^2\log\left(1-\frac{r^2}{r_{\text{max}}
^2} \right).
\end{equation}
Here $k$ is the spring constant and $r_ {\text{max}} $ is the maximum
allowed separation between the neighbouring polymer monomers. Excluded-volume
interactions between polymer segments  are given by the standard truncated Lennard-Jones
(LJ) repulsive potential (Weeks-Chandler-Andersen potential),

\begin{equation}
U_{\text{LJ}}(r,\epsilon)=\left\{\begin{array}{ll}
4\epsilon[(\sigma/r)^{12} - (\sigma/r)^{6}] + \epsilon,& r< r_{\text{cutoff}}\\
0, & \text{otherwise}
\end{array}
\right.
\label{eq-lj}
\end{equation}
with $r_{\text{cutoff}}=2^{1/6}\sigma$.
Here, $r$ is the monomer-monomer distance, $\sigma$ is the chain monomer
diameter, and $\epsilon$ is the strength of the potential. We set  $k = 30
$, $r_ {\text{max}} = 1.5$ (to minimise bond crossings \cite{kremer-grest} of the chain), and $\epsilon = 1$ (with all the energies being
measured in units of the thermal energy, $k_\text{B}T$). Similar repulsive 6-12
LJ potentials parameterise the (chain monomer)-crowder and crowder-crowder
interactions. 

The chain monomer diameter is set in simulations to $\sigma = 4  $
nm, determining polymer thickness and its effective viscosity in the crowded
solution, $\eta$. The diameter  $d_{\text{cr}}$ of mono-disperse hard-core repulsive crowding particles varies in simulations in the range $0.75 \leq d_{\text{cr}}\leq 8\sigma$. The mass density is kept constant for all crowder sizes, fixed to the value known for average cytoplasm-crowding macromolecules \cite{loop-crowd-simul-1}. Thus, for the varying crowder sizes its mass grows as $m_{\text{cr}}\sim d_{\text{cr}}^3$ and the friction coefficient increases according to the "effective" Stokes-Einstein law as $\xi_{\text{cr}}\sim d_\text{cr}$, similar to the procedure of Ref. \cite{kaifu13}. We use a cubic simulation box with volume $V=L^3$ and periodic boundary conditions. The volume fraction of crowders is $\phi=N_{\text{cr}}V_{\text{cr}}/V$, where $N_{\text{cr}}$ is the number of crowders and $V_{\text{cr}}=\frac{4}{3}\pi(d_{\text{cr}}/2)^3$ the volume of each crowding particle. The characteristic time scale for a crowder with $d_\text{cr}=1\sigma$ and $m_{\text{cr}}$=67.7 kDa \cite{loop-crowd-simul-1} is $\delta\tau=d_{\text{cr}}\sqrt{m_{\text{cr}}/(k_\text{B}T)}\approx$ 0.36 ns. The times presented in the figures below are in the units of this elementary time step $\delta\tau$. The features of the crowder size we observe with this explicit simulation scheme would not be visible in more 
coarse-grained models of crowded media employed previously, including those with effective 
depletion potentials.  

The dynamics of position $\mathbf{r}_{i}(t)$ of the chain monomers  is described by the Langevin equation
\begin{eqnarray}
\nonumber
m\frac{d^2\mathbf{r}_i(t)}{dt^2} =\\ 
\nonumber
&&\hspace*{-1cm}
-\sum_{j=1,j \neq i}^{N}\boldsymbol{\nabla} U_ {\text{LJ}} (|\mathbf{r}_i-\mathbf{r}_j|) -\boldsymbol{\nabla} U_ {\text{FENE}} (|\mathbf{r}_i-\mathbf{r}_{i\pm1}|) \\
&&\hspace*{-1cm}
-\sum_{j=1}^{N_{\text{cr}}}\boldsymbol{\nabla} U_ {\text{LJ}} (|\mathbf{r}_i-\mathbf{r}_{\text{cr},j}|)- \xi\mathbf{v}_{i} (t)+ \mathbf{F}_i(t). 
\end{eqnarray} 
Here $m$ is the mass of the monomer, $\xi$ is the monomer friction
coefficient and $\mathbf{F}_i(t)$ is the white Gaussian noise with the
correlator $\left< \mathbf{F}_{i}(t)\cdot \mathbf{F}_{j} (t') \right>= 6\xi
k_\text{B}T\delta_{ij}\delta (t - t') $ that couples the particle friction
and diffusivity $D=k_\text{B}T/\xi$. Similarly to the procedure described in
Ref. \cite{shin14melting}, we implement the velocity Verlet algorithm with
the integration time step of  $0.002 \leq \Delta t \leq  0.01$. Smaller simulation
step was used for bigger crowders and higher volume fractions $\phi$. 

The
terminal monomers interact with the energy $\epsilon_s$ which mimics e.g. the
energetic profit for the formation of closed ssDNA hairpin structures via
hydrogen-bonding pairing interactions between the complementary bases on the
end DNA fragments. Although we simulate flexible polymers, via corresponding
rescaling the effective monomer size, the results can be applicable to
looping of semi-flexible dsDNA as well, where the loop/ring joining reaction is
often supported by the ligation enzymes \cite{langowski-ligation,volo-ligation}.
The number of the chain monomers $n$ vary in simulations in the range
$10 \leq n \leq 256$. The pairing energy of $\epsilon_s = 5 k_\text{B}T$ used in
the majority of results below can be considered as a good estimate
for pairing propensity in DNA hairpins with not too long complementary
ends, see Ref.~\cite{weiss13crowd}. We examine the range of chain end
cohesiveness of $0 \leq  \epsilon_s  \leq 10 k_\text{B}T$. 

We simulate the attractive
end-to-end interactions via the same LJ potential, Eq.~\eqref{eq-lj},
but with larger cutoff distance and bond intensity $\epsilon_s$, namely
$U_{\textrm{attr}}(r)=U_{\text{LJ}}(r,\epsilon_s)+C_{\text{LJ}}$ and $r_{\text{cutoff}}=3\sigma$. Along with this longer cutoff distance we shift the entire LJ potential in the vertical direction by the constant $C_{\text{LJ}}$ so that at $r=r_\textrm{cutoff}$ the potential becomes continuous with the zero-value branch at larger distances $r>r_{\text{cutoff}}$. The volume fraction of mono-disperse crowders is varied in our simulations up to $\phi=0.3$; see Ref. \cite{greek-dense} for even denser colloidal systems.

The free energy of ssDNA hairpin formation contains two contributions: the
favourable stacking/pairing of the helical dsDNA part and the entropic penalty
of the looped part. The sum of the two for real DNAs is a complicated function of DNA
sequence and other model parameters \cite{jost09,santalucia98} amounting to $\sim$-2.1 kcal/mol$\approx$-3.5 $k_\text{B}T$ for about 20 bp long DNA hairpins used in
Ref. \cite{weiss13crowd}. Longer complementary paired stem parts result in
more stable hairpins, we mimic in simulations via larger values of end-to-end
cohesive energy $\epsilon_s$.

We neglect long-range interactions between polymer segments, including the
electrostatic forces, that is a reasonable approximation for long chains
at physiological salt concentrations. In low-salt solutions, however, in
application to DNA, the charge-charge electrostatic interactions will become
important for the loop-closure probability and dynamics \cite{ac11rings2}. We
assume polymer-solvent interactions stays unaltered at increasing volume
occupancies by crowders (see Ref. \cite{japan13CR-NAcrowding} for possible
effects of MMC onto the properties of nucleic acid solutions at reduced
solvent activity).

The  hydrodynamic interactions are also neglected below (the Rouse polymer
model), see Refs. \cite{skolnick-hdi-mmc,winkler-rouse-relaxation,
winkler-rouse-relaxation-fcs} for some implications. The effects
of hydrodynamic interactions onto end-monomers dynamics of dsDNA has
been studied by fluorescence correlation spectroscopy experimentally in
Refs. \cite{krich04,petrov06}. Theoretically, the Rouse versus Zimm chain dynamics
has been examined for semi-flexible polymers in  solutions \cite{polymers-referee-solution-1, polymers-referee-solution-2}, confined spaces \cite{polymers-referee-confined}, and near surfaces \cite{hinz-surface}. In the latter situation e.g. it was clearly demonstrated,
based on the hydrodynamic Brownian simulations and the mean-field hydrodynamic
theory, how the Zimm dynamics turns into the Rouse one as the polymer
chain approaches the no-slip surface \cite{hinz-surface}. In particular,
for the end-to-end distance of the chain near the interface, the influence of
hydrodynamic interactions screened as $\propto~1/r$ with the inter-particle distance,
was shown to be marginal. \footnote{In highly-crowded systems, which are the main targets of the current study,
the polymer chain experiences collisions with many crowders around in the
course of diffusion-limited looping. We thus believe  hydrodynamic
interactions to be of secondary importance for the static and dynamical
effects considered here, likely just re-normalising the effective viscosity of the solution. As we demonstrate, rather the size of thermally-agitated crowders, which are to be displaced to ensure polymer
looping, and their volume fraction are the dominant effects.}

\subsection{Parameters and Data Analysis}

We compute the end-joining statistics from the time series of the polymer
end-to-end distance generated in simulations as follows. For looping, we
start with the 
most probable end-to-end chain extension (the
minimum of the free energy $F(r)$, see Eq. (\ref{eq-free-energy}) and Fig.~\ref{fig4} below,
$r=r_{\text{eq}}$) and let the chain ends diffuse to the final extension $r=r_f\approx
1.2 \sigma$. (The contact distance between the terminal chain beads in the
folded state implemented in Ref. \cite{davide06crowd} was somewhat different,
$r_f=\sigma+d_{\text{cr}}$.) The looped state distance $r_f$ corresponds to the
minimum of the LJ potential in Eq.~\eqref{eq-lj} and stays nearly constant
in the whole range of model parameters used here.

The time required for the chain to join its ends is defined as the looping
time $T_l$, see Fig. \ref{fig2}. The unlooping time $T_{ul}$ is defined as the time required for the chain
to expand back, from the jointed-ends state with $r=r_f$ to the equilibrium
state at  $r=r_{\text{eq}}$. This distance is a function of all model parameters,
in particular of the chain length $l=n\sigma$ and the MMC fraction, that is
accounted for in simulations below. The closing time $T_{cl}$ is defined as the
average time the polymer  needs to diffuse from the last moment its end-to-end
extension was   $r=r_{\text{eq}}$ to the first moment with the close-contact distance
of $r\approx r_f$. The opening time $T_{op}$ is the minimal time for the chain
ends to diffuse from the closed state $r=r_f$ to a first state with $r=r_{\text{eq}}$. In
Fig. \ref{fig2} we illustrate on a real end-to-end diffusion
trace the definitions of the looping/unlooping and opening/closing times.

Likewise, the critical distance of $r_c=1.75 \sigma$ used below to define the
occurrence of end-monomer contacts stays nearly constant. It approximately
denotes the end-monomer separation at which the free energy barrier emerges
which separates the close-looped and equilibrium states of the polymer,
see Fig.~\ref{fig4} below. This critical distance $r_c$ is used
below to compute the looping probability $P_l$. One can think of other
choices for $r_c$ to mimic somewhat longer-ranged nature of end-end contacts.
\footnote{Note that starting from randomised chain configurations, in simulations
of Ref. \cite{davide06crowd} the looping time was computed till the chain
ends are closer than a "critical" distance $r_c$. The latter is an important
parameter that depends on the type of interactions which act between the
chain ends. It has a meaning of effective inter-segmental distance at which
e.g. DNA-protein-DNA contacts can be established, $\approx$ 3-5 nm for a
typical transcription factor.}

We study the end-to-end joining statistics; the implications
of MMC onto looping kinetics of inner polymer monomers is beyond the scope
of this study and will be presented elsewhere. The simulation time for the
chains of $n$=8, 32, and 128 monomers on a standard 3-3.5 GHz core machine
is about 3, 4, and 60 h, respectively. The typical number of the looping
events used for averaging procedure for these chain lengths is about 2000,
500, and 200, correspondingly. In some cases we use traces, that
are twice as long, for a better statistics. The number of crowding molecules of size $d_{\text{cr}}=1\sigma$ in the simulation box used
to perform simulations of the polymer chains of these lengths is $N_{\text{cr}}\approx
1000$, $3000$, and $10000$, respectively. Moreover, we remark that instead of averaging over the ensemble of initial chain
configurations, we rather analyse the individual simulated time traces
of the end-to-end distance $r(t)$ to compute the chain looping
characteristics.

We analysed the $r(t)$ data obtained from either single or multiple simulation runs, depending on the total computation time used. The typical running time,  $t\sim 10^{5\dots 7}\times {\delta \tau}$, is chosen much longer than all the time scales in the system, in order to avoid a bias in sampling of end-joining events. To perform the error analysis, we use different methods for the dynamic and static quantities. As looping events are rare, the time intervals between them are of the order of the chain relaxation time, and the events can be considered independent. Thus, we use the standard error of the mean to compute the error bars for the looping (unlooping) and opening (closing) times. For the static quantities, such as the radius of gyration of the polymer, we split the entire trajectory into ten sub-series, calculate the values for each of them, and then compute the standard deviations of those pre-averaged values to get the final error bar. Previously \cite{shin14rings} we also used the so-called "blocking method" for the error analysis in correlated sets of data. Here we compare the two methods for a number of quantities and the differences in the sizes of the error bar were $\lesssim 30\%$.

We need to distinguish the MMC effects for small  ($d_{\text{cr}} \ll
R_g=\sqrt{\left<R_g^2\right>}$) and large crowders ($d_{\text{cr}} \gtrsim
R_g$). Large crowders creates voids/cages between themselves which facilitate
compaction of relatively short polymers and facilitate looping. The reader is referred to Sec. \ref{sec-caging} for the quantitative analysis of caging effects in our polymer-crowder mixtures.  For
longer chains, which do not fit into a single cavity and need to occupy
the neighbouring voids, the effect of crowders on looping probability can
be \textit{inverted}. A similar effect occurs in MMC-mediated protein folding, when
small crowders favour the compact state of a protein, while larger ones can
promote protein unfolding  \cite{sschreiber}. The systematic investigation of
crowder  surface properties is the subject
of our future investigations \cite{looping-future}.

\section{Results: Crowding and Polymer Dynamics}
\label{sec-results}

\begin{figure}
\begin{center}
\includegraphics[width=7cm]{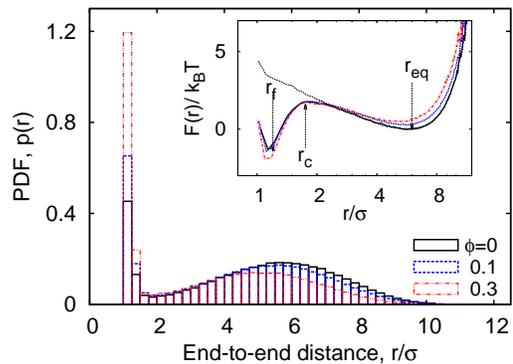}
\end{center}
\caption{Bimodal distribution $p(r)$ of the polymer end-to-end distance at varying
MMC fraction $\phi$. The inset is the free energy profile for looping, $F(r)$, with the most likely separation between the polymer ends shown as $r_{\text{eq}}$. The free energy profile for purely repulsive end monomers $(\epsilon_s=0$, $\phi=0$) is also shown as the dotted curve in the inset. Parameters: $\epsilon_s=5k_\text{B}T$, $n=16$, $d_{\text{cr}}=1\sigma$.
\label{fig4}}
\end{figure}

\begin{figure*}
\begin{center}
\includegraphics[height=4cm]{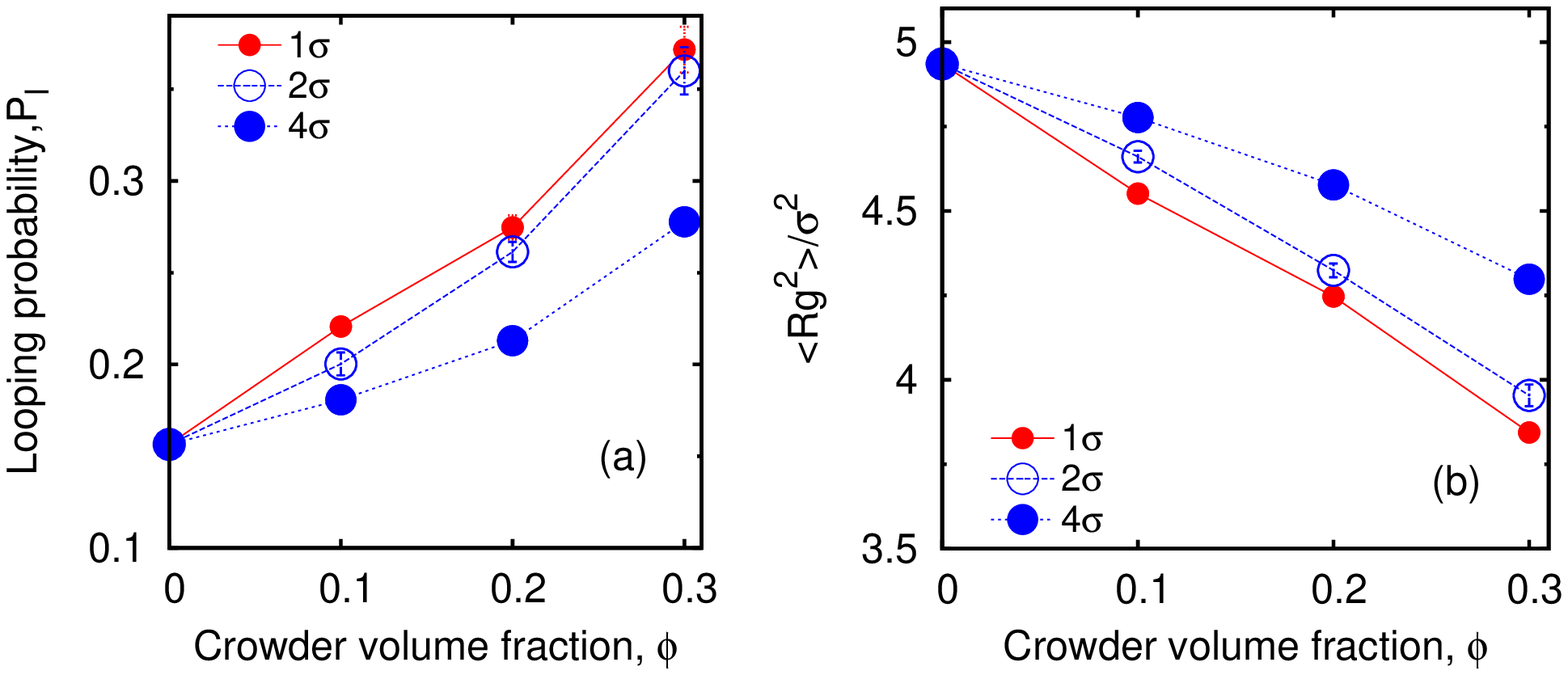}
\hspace*{0.2cm}
\includegraphics[height=3.8cm]{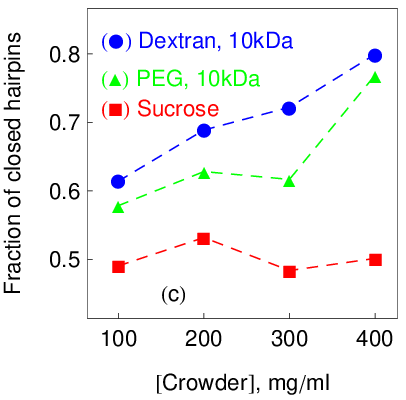}
\end{center}
\caption{Looping probability $P_l(\phi)$ (panel a) and polymer radius of
gyration $R_g(\phi)$ (panel b) computed from simulations for $\epsilon_s=5k_\text{B}T$, $n=16$. Experimental data \cite{weiss13crowd} 
for the fraction of closed ssDNA hairpins is shown in panel (c). Here and below the simulation data for $d_{\text{cr}}=1\times\sigma$ are presented as small red circles, big blue circles correspond to  $d_{\text{cr}}=4\times\sigma$. Note that the weaker effect of larger ($d_{\text{cr}}=4\times\sigma$) crowders onto the dimensions of the polymer coil shown in panel (b) may be due to the matching of sizes ($R_g\sim d_{\text{cr}}$) for the relatively short chains considered here.
\label{figs1}}
\end{figure*}

The equilibrium statistical behaviour of a linear flexible polymers with
sticky ends is governed by the tradeoff between the enthalpically favourable
pairing of the sticky ends and the entropy loss in the more compact looped
state. In what follows we first rationalise the effects of MMC on the static
properties of polymer looping. We then examine the kinetics of loop closure
and opening as functions of the details of the crowders such as crowding fraction
and crowder size.

\subsection{Distribution Function $p(r)$ and Free Energy}

Our simulations generate time traces of the end-to-end distance $r(t)$
between the two extremities of the linear polymer. These two end monomers interact
through an attractive LJ potential with an attractive cohesiveness $\epsilon_s$ which is
varied in the range $0\leq\epsilon_s\leq 10 k_\text{B}T$, see the specification of the system in the
preceding section. The recorded dynamics for $r(t)$ exhibits
the highly erratic dynamics shown in Fig.~\ref{fig2}, see below for the exact
definition of the looping and unlooping times. We first focus on the one-dimensional probability
density function (PDF) $p(r)$ of the end-to-end distance, as shown in
Fig.~\ref{fig4}. 

The relative motion of terminal monomers is subject to the free energy
potential $F(r)$ that can be obtained from the PDF of the end-to-end distance  $p(r)$ (see Fig. \ref{fig4}) via the inverse
Boltzmann relation as \begin{equation}F(r)=-k_\text{B}T\log[p(r)].\label{eq-free-energy}\end{equation}
The presence of sticky chain ends gives rise to the formation of a double-well
 potential for $F(r)$, see Figs. \ref{fig4} and \ref{figs2}. The shallow free energy well related to the maximum of the PDF $p(r)$ corresponds to the equilibrium end-to-end chain distance in the absence of sticky ends, namely $r=r_{\text{eq}}$. This minimum is
 accompanied by a sharp free energy well at very close end-to-end distances
 due to the presence of sticky ends. The transition between the looped
and unlooped states of the polymer takes place in this asymmetric $F(r)$
 potential. The chain should overcome free energy barriers in the course
of looping and unlooping. Simultaneously, the equilibrium chain extension
$r_{\text{eq}}(n)$ is a growing function of the chain length, see Fig. \ref{figs2}. \footnote{Note that the one-dimensional end-monomer distribution function $p(r)$ does not involve a Jacobian to recover the free energy profile $F(r)$ since our end-to-end distance measurements already account for the spatial dilation.}

\subsection{Looping probability and polymer size}

From $p(r)$---which is a function of the number of monomers $n$---we
compute the probability distribution \begin{equation}P_l=\int_{\sigma}^{r_c}p(r)dr\end{equation} for the chain to be in the looped state as function of $n$, see Fig. \ref{figs2}. That is, $P_l$ is proportional
to the number of configurations in which the sticky ends of the chain are within a
maximum distance of $r_c=1.75\sigma$. The lower cutoff discards thermodynamically
unfavourable, rare events when the end beads are closer than the distance $\sigma$.
For a single trajectory $r(t)$ of the end-to-end distance shown in
Fig.~\ref{fig2}, the probability $P_l(n)$ is then equal to the fraction of time
during which the chain is looped.

\begin{figure}
\includegraphics[width=8.5cm]{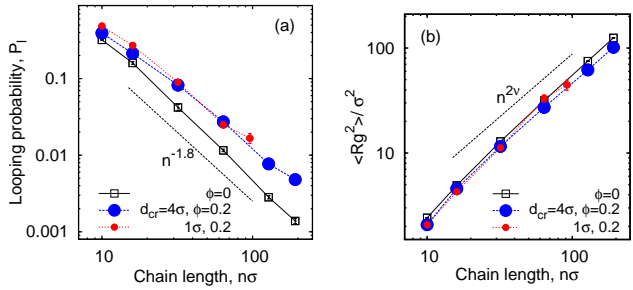}
\caption{Looping probability $P_l$ and the gyration radius $R_g$ versus the
degree of polymerisation $n$. The asymptotes $P_l(n)\sim n^{-1.8}$ and $\left<R_g^2
(n)\right>\sim n^{2\nu}$ correspond to the dashed lines. Parameters: $\epsilon_s=
5k_\text{B}T$ and $\phi=0,0.2$. The crowder sizes are as indicated.
\label{fig3}}
\end{figure}

Fig.~\ref{figs1}a demonstrates that the looping probability $P_l$ grows with the
crowding fraction $\phi$. This is in accord with recent results of ssRNA tertiary folding-unfolding dynamics \cite{dupuis} as well as ssDNA hairpin
formation measurements \cite{weiss13crowd} in crowded polymeric solutions.
In the latter experiment, fluorescence correlation spectroscopy data indicated a
linear increase of the fraction of closed ssDNA hairpins as function of $\phi$,
\begin{equation}P_l(\phi)\sim A+B\phi,\label{eq-p-phi}\end{equation} see Fig.~\ref{figs1}c. Our results reported here demonstrate
that this trend becomes amplified for growing length $n\sigma$ of the polymer, as
demonstrated in Fig.~\ref{fig3}a. The magnitude of the relative facilitation for
the looping probability for $\phi\approx0.2$ is of the order of 2 to 4, compared
with the dynamics in the absence of crowders. This value is similar to the
experimental trends for ssDNA hairpin formation with MMC \cite{weiss13crowd},
compare Fig.~\ref{figs1}c. This $P_l-$enhancement effect is present for both small and large
crowders, as shown in Fig. \ref{fig3}a.

Consider now the PDF $p(r)$ shown in Fig.~\ref{fig4}. It has a bimodal structure,
reflecting the proximity between the sticky ends with end-to-end distances $r
\approx\sigma$ and a broad distribution of $r$ values reflecting the diffusive
nature of the chains ends in the extended state. As can be seen in Fig.~\ref{fig4},
the presence of MMC favours more compact polymer states: with increasing crowding
fraction $\phi$ the polymer radius of gyration $R_g$ decreases, in accord with
common MMC effects \cite{minton08,protein-folding-crowd-confin-review}. The
significant shift of the distribution to shorter $r$ values is particularly
visible when the peak around $r\approx\sigma$ is considered. 

For completeness we
mention that, as expected a priori, stronger cohesiveness of the sticky ends
favours higher looping probabilities $P_l$, reaching unity at $\epsilon_s\gg1k_BT$
(see Fig.~\ref{figs4}) and yields progressively longer unlooping times. This fact also agrees with the experimental data on ssDNA
hairpin formation in solutions of polymeric crowders of different MWs
shown in Fig.~\ref{fig7}. Interestingly, we find that for crowder molecules with
larger diameter $d_{cr}$ the looping probability $P_l$ becomes less sensitive to
$\phi$, as demonstrated in Fig.~\ref{figs1}a (smaller values of $B$ in Eq. (\ref{eq-p-phi})). To map the
detailed parametric dependence of the looping statistics as function of chain
length $n$, crowder size $d_{cr}$, and fraction $\phi$ is a major challenge for
simulations. We examine all these effects below.

As shown in Fig.~\ref{fig3} for both small and large crowders the looping
probability $P_l$ decreases with the chain length $n$ as
\begin{equation}
P_l(n)\sim n^{-1.8}.
\end{equation}
The scaling exponent $1.8$ law is close to the one of the Stockmayer formula
$P_l(n)\simeq n^{-3\nu}$ for the looping of a self-avoiding polymer, where $3\nu
\approx1.76$ \cite{grosberg}. At the same time the radius of gyration of the chain
grows as $\left<R_g^2(n)\right>\simeq n^{2\nu}$, as expected for a self-avoiding
chain \cite{grosberg}. These dependencies are seen to be quite generic for varying
crowder sizes $d_{cr}$ and fractions $\phi$, see Fig. \ref{fig3}.

The simulations yield instructive shapes for the polymer free energy $F(r)=-k_\text{B}T
\log[p(r)]$ as shown in the insets of
Fig.~\ref{fig4} and in Fig.~\ref{figs2}. We find a clear trend for the free energy
barriers $\Delta F(n)$: for the transition from the unlooped to the looped state
the barriers become higher for longer chains, as shown in Fig.~\ref{figs2}. This
effect is due to the higher entropic penalty upon looping for longer polymers. In
contrast, the barriers for a transition from the looped to the unlooped state are
fairly insensitive to $n$, reflecting that unlooping is a \textit{local activation} effect
of dissolving the bond between the terminal monomers. This important feature gives
rise to a more pronounced chain-length effect on the looping time $T_l$ as compared
to the analogous dependence of the unlooping time $T_{ul}$, as seen in
Fig.~\ref{fig5}. We now study the (un)looping times in more detail.

\subsection{Looping and unlooping times}

Fig.~\ref{fig2} shows how we extract the average looping and unlooping times $T_l$ and
$T_{ul}$ from the time series $r(t)$ of the end-to-end distance. Namely, $T_l$
is counted from the point when---after a previous looped state---the chain ends
reach their equilibrium distance $r_{\text{eq}}$ until they touch close to the minimum of the
attractive LJ potential. By definition, the equilibrium distance $r_{\text{eq}}$
corresponds to the free energy minimum for the extended chain conformations.
From that moment, $T_{ul}$ is counted until the chain ends are separated by the
distance $r_{\text{eq}}$ again. The computation of $T_l$ thus involves extensive chain rearrangements and thus non-trivially depends on the crowder fraction
$\phi$, which favours more compact states. In our analysis $T_l$ and $T_{ul}$ are
then averaged over many looping events, the results being shown in Fig.~\ref{figs3}
for a fixed chain length.

The distribution of looping times is found to be nearly
exponential, and the characteristic time is shorter in more crowded solutions
of bigger crowders, see Fig.~\ref{figs5}. The full statistics and fluctuations of  $T_l$ can be envisaged from the
PDFs presented in Fig. \ref{figs5}. We fitted the $p(T_l)$ functions by
two-parametric Weibull distributions of the form \begin{equation}p(T_l) \sim
T_l^{\gamma-1}\exp [-(T_l/T_l^\star)^\gamma].\label{eq-weibull}\end{equation}
We found that the looping times are nearly exponentially distributed, with the
parameter $1 \lesssim  \gamma \lesssim 1.17 $ being quite close to unity for all
$\phi$ fractions and crowder sizes examined in Fig. \ref{figs5}. Note that the nearly---but not exactly---exponential distribution $p(T_l)$ is indicative of some short-living "intermediates" in the looping process. Note also that the first-encounter kinetics of the polymer ends is reminiscent of the first-passage kinetics of reactants in generalised biochemical networks, see e.g. Refs.  \cite{tolja1,tolja2}. The decay
length $T_l^\star$ of $p(T_l)$ distributions appears to be growing with $\phi$ for small crowders,
while the decay of $p(T_l)$ gets faster with  $\phi$ for larger crowders,
see Fig. \ref{figs5} for $d_{\text{cr}}=1\sigma$ and $4\sigma$. This behaviour is
physically consistent with the more restricted motions of the whole polymer
and its ends at higher MMC fractions of bigger obstacles: the looping kinetics
becomes faster and the spread of looping times gets  narrower (more reliable
looping events).

We also consider the opening and closing times $T_{op}$ and $T_{cl}$
(Fig.~\ref{fig2}). $T_{op}$ is the time for the chain ends to open up from a
closed state and first reach the equilibrium distance $r_{\text{eq}}$.
$T_{cl}$ measures the time from the last occurrence of $r_{\text{eq}}$ before a new
looping event with $r<r_c$. Both $T_{op}$ and $T_{cl}$ grow with $\phi$, as shown
in Fig.~\ref{figs6}. These times are, as expected, much shorter than the looping
and unlooping times. For $T_{op}$ and $T_{cl}$ we detect no significant difference
in their $\phi$ dependence, consistent with theoretical \cite{hummer2004} and
experimental \cite{neupane2012} results.

\begin{figure}
\includegraphics[width=8.8cm]{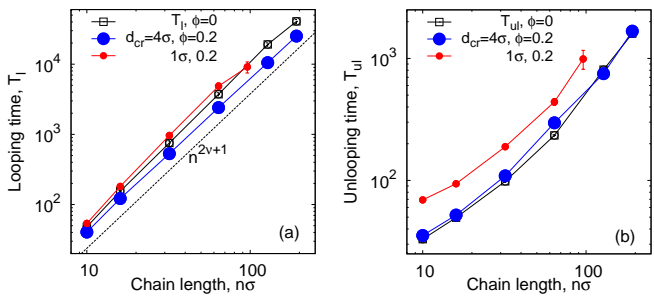}
\caption{Average looping (panel a) and unlooping (panel b) times versus the chain length. The
asymptote (\ref{eq-tl}) of  $T_l(n)\sim n^{2\nu+1}$ in absence of crowders is
shown by the dashed line in panel (a). Parameters are the same as in Fig. \ref{fig3}.
The shown error bars are often smaller than the symbol size.
\label{fig5}}
\end{figure}

So what about the dependence on the crowder size? Fig.~\ref{fig5} demonstrates
that for small crowders the looping kinetics is somewhat inhibited and $T_l$ increases with
$\phi$. For large crowders, however, we observe the opposite and stronger trend: polymer looping
is facilitated. As detailed in Fig.~\ref{figs3}a, $T_l$ indeed decreases with $\phi$
up to $d_{\text{cr}}=4\sigma$, however, for even larger crowders it starts to
increase again, see Fig. \ref{figs3}b. For very large crowders $T_l$ appears to approach the looping
time in absence of crowders, indicated by the dashed line in Fig.~\ref{figs3}b. Fig.~\ref{figs7} reveals that the solution viscosity increases more
strongly with $\phi$ for small crowders, slowing down the chain dynamics
and reducing the looping rates. This non-trivial behaviour illustrated in Fig.~\ref{fig5} is our \emph{first key
result}.

Apart from the viscosity dependence, in Fig.~\ref{fig6} we highlight another important
crowding-mediated effect. Namely, when the crowders are small, entropic effects favour a good mixing of crowders and chain monomers with little implications of the chain connectivity. When the crowders become larger,
however, depletion effects become increasingly dominant. The chain becomes confined in a ``cage''. We emphasise here that the cage is not static but rather a dynamic entity, because of perpetual diffusion of crowders. Only at very high $\phi$ values or with possible attractions between the crowders the cage becomes static, as studied in Ref. \cite{poon09pnas}. In this confined state, the looping probability is significantly
increased and thus the looping dynamics gets facilitated. As shown in Fig.~\ref{fig5}
the depletion effect just outweighs the increased viscosity for larger crowders. The dynamics of crowders remains Brownian even at high volume fractions of $\phi\sim0.3$, see below.

The unlooping time $T_{ul}$, in contrast, typically increases with $\phi$. As shown
in Figs.~\ref{fig5} and \ref{figs3}c, while the dependence of $T_{ul}$ on $\phi$
is very weak for large crowders, it becomes quite sizable for smaller crowders.
The effect on $T_{ul}$ is due to both the higher
viscosity induced by MMC and the impeded chain opening imposed by the caging
effects. For the unlooping process both effects do not lead to an
inversion of the $\phi$-dependence of $T_{ul}$ inhibiting chain opening. The
unlooping time is a monotonically decreasing function of the crowder size, see
Fig. \ref{figs3}d.\footnote{The unlooping time is much shorter than the looping time and $T_{ul}$ shows a weaker growth with $n$ than $T_l$, compare the two panels in Fig. \ref{fig5}. One possible reason is as follows. A looping event is the end-monomer encounter reaction that becomes progressively slower for larger polymer coils. The unlooping time in contrast  is related to the (only moderately perturbed) diffusion of the polymer ends on the length-scale of the polymer coil, $r_{\text{eq}}(n)\sim R_g(n)$.}

\subsection{Comparison with DNA hairpin formation experiments}

\begin{figure}\begin{center}
\includegraphics[height=4cm]{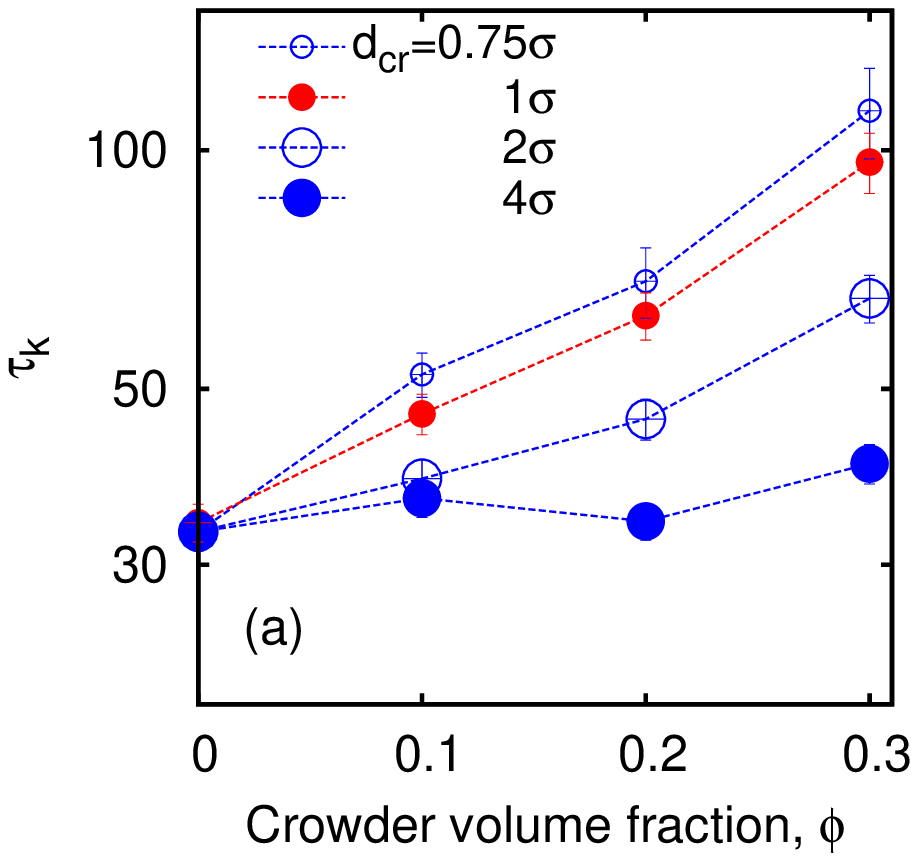}
\includegraphics[height=4cm]{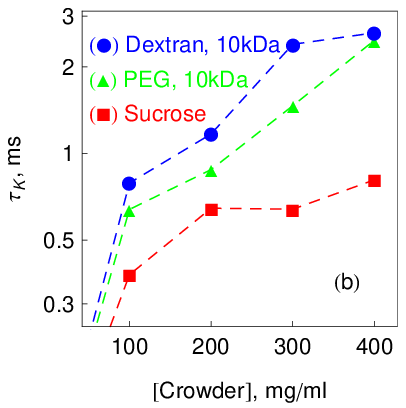}
\end{center}\caption{Characteristic time $\tau_K$ given by Eq. (\ref{eq-tauK}) and computed from simulations (a) for $n=16$
and $\epsilon_s=5k_\text{B}T$. (b) Experimental data \cite{weiss13crowd} for ssDNA hairpin formation kinetics in solutions of different crowders. Both graphs are in
the log-linear scale.
\label{fig7}} \end{figure}

Ref.~\cite{weiss13crowd} reports experimental data from fluorescence correlation
measurements of ssDNA hairpin formation. The characteristic time $\tau_K$ for
the measured fluorescent blinking is given by the harmonic mean \cite{weiss13crowd},
\begin{equation}
\tau_K=T_l T_{ul}/(T_l + T_{ul}).
\label{eq-tauK}
\end{equation}
Similar to the experimental data \cite{weiss13crowd}, we show that $\tau_K(\phi)$ has a tendency to grow with $\phi$ for crowders of all sizes and polymers of all lengths examined in the simulations, see Figs. \ref{fig7} and \ref{figs7a}.
We observe that the typical variation of $\tau_K$ with $\phi$ corresponds to a
factor of 2-3, in agreement with the measured data \cite{weiss13crowd}, as shown
in Fig.~\ref{fig7}. We also reveal a systematic dependence of the crowder diameter onto $\tau_K$ enhancement, in which smaller crowders are most efficient, see Fig. \ref{fig7}. The curves in the plots indicate a nearly exponential dependence
\begin{equation}\tau_K(\phi)\simeq\exp(\gamma\phi)\end{equation} as function of the crowding fraction $\phi$.
This is consistent with the exponential dependence of the self-diffusivity of a
tracer in crowded solutions, $D(\phi)\sim\exp(-\gamma \phi)$ \cite{minton82}. We
checked that looping of \textit{longer} polymers in crowded solutions yield qualitatively
similar enhancement effects on $\tau_K$ with $\phi$, see Fig.~\ref{figs7a}. In this figure the crowders are fairly large, $d_{\text{cr}}=4\times\sigma$, and the magnitude of $\tau_K$ enhancement is somewhat smaller, consistent with the behaviour of $\tau_K(d_{\text{cr}})$ presented in Fig. \ref{fig7}a.

To make the quantitative comparison of our results for $\tau_K$ to the
experimentally observed $\tau_K(\phi)$ enhancement \cite{weiss13crowd},
one needs to compare the relative sizes of polymers and crowders (experiment versus
simulations). Namely, 10 kDa PEG polymers have $R_{g,\textrm{PEG}}\approx$2.8 nm,
while for 21-bp long DNA hairpins  $R_{g}\sim$7 nm \cite{weiss13crowd}. In
simulations, for $n=16$ chains, see Fig. \ref{figs3}, the gyration radius
is $R_{g}\approx  2.5\sigma$ (Fig. \ref{figs1}b), so the crowders of
diameter $d_{\text{cr}}\approx 2\sigma$ are in the same relation to the polymer
size in simulations as 21-bp DNA hairpins to 10 kDa PEG in experiments
\cite{weiss13crowd}. Note that 10 kDa branched dextran polymers are
considerably smaller than 10 kDa PEG \cite{weiss13crowd} and the dynamics
of DNA hairpin formation is slower in dextran solutions. The physical
reason for this behaviour, as proposed in Ref. \cite{weiss13crowd}, is a
pronounced sub-diffusion of DNA hairpins in solutions of dextran, with the
scaling exponent of $0.7<\beta<0.85$, in stark contrast to the sucrose and
PEG solutions where the hairpin diffusion is nearly Brownian ($
0.9<\beta<1$), see also below.

\subsection{Length dependence and effective diffusivity}

The observed "tug-of-war" between facilitation and inhibition is a fundamental
feature of the looping kinetics for all chain lengths. Fig.~\ref{fig5}
illustrates that looping is systematically facilitated for larger crowders and
impeded for smaller crowders. Concurrently, the scaling of the looping time with
$n$ given by Eq.~\eqref{eq-tl} \textit{does not} change appreciably in crowded
solutions compared to the dilute case $\phi=0$, as shown in Fig.~\ref{fig5}. This
is our \emph{second important result}. 

Note that for longer polymers the accessible space inside the coil increases and at some point even large crowders can be accommodated therein, thus reverting effect of polymer compaction by MMC. However, the gyration radius of our longest chains with $n \approx 200$ monomers is still too small to see this happen for the larger crowders ($d_{\text{cr}}=4\times\sigma$) studied. Thus, the $T_l(n)$ scaling behaviour for even longer chains remains similar to the situation in absence of crowders.

We observe a slightly more pronounced
looping time variation with $\phi$ for longer polymers in crowded solutions, in
agreement with well-established results, for instance, in protein-DNA interactions
\cite{leduc13}. The unlooping times vary substantially with the polymer length (in
stark contrast to the observations of Ref.~\cite{davide06crowd}). This indicates
that the unlooping is not a purely local unbinding process, but it needs the
cooperative motion of the polymer. \footnote{Such a statement is valid for \textit{relatively weak} cohesion strength of the terminal monomers. In contrast, for very large end-to-end binding energies the unbinding kinetics is dominated by the dynamics of terminal monomers only, as illustrated by the Arrhenius-like behaviour for the unbinding events in Fig. \ref{fig9} below. The motion of the polymer chain enables the accumulation of the energies  $\gtrsim k_\text{B}T$ required to disrupt the bond between the polymer ends. }

\begin{figure}
\begin{center}
\includegraphics[width=7cm]{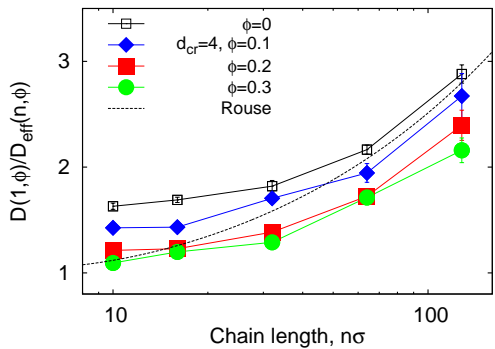}
\end{center}
\caption{Reciprocal effective diffusivity $D(1,\phi)/D_{\mathrm{eff}}(n,\phi)$ of
polymer ends, obtained from fit of the $T_l$ data in Fig.~\ref{fig5} with
Eq.~\ref{eq-tl-fe-shape}. The Rouse chain result $D_{R}(n,0)$ given by Eq. (\ref{eq-thirum-diff}) with $D(1,0)=1/2$
as used in simulations is the dashed curve.
\label{fig8}}
\end{figure}

In Fig.~\ref{fig8} we study how many chain monomers are involved in looping
events by quantifying the inverse \textit{effective\/} position-independent diffusivity $1/D_{
\mathrm{eff}}$ of the end monomers. We
use the data of Fig.~\ref{fig5} for $T_l$ and the general expression for
mean first-passage (i.e., looping) times
\begin{equation}
T_l=\int_{r_c}^{r_{\text{eq}}}dr'\frac{e^{F(r')/(k_\text{B}T)}}{D_{\mathrm{eff}}}
\int_{r'}^{n\sigma}dr''e^{-F(r'')/(k_\text{B}T)},
\label{eq-tl-fe-shape}
\end{equation} 
in a general potential $F(r)$ \cite{kampen}. Here $n\sigma$ is the maximal chain
extension. We fit the simulation data for $T_l$ with the free energy profiles
$F(r)$ computed for each chain length in Fig.~\ref{figs2}. The effective
end-to-end diffusivity in the model of Rouse chains without crowding as derived in Ref. \cite{loop-simul-3}, \begin{equation}D_{R}(n,0)
/D(1,0)\approx8/\sqrt{\pi n}-16/(3n),\label{eq-thirum-diff}\end{equation}  is represented by the
dashed line in Fig.~\ref{fig8}. Although our simulation data in the limit $n\gg1$
follow this Rouse-chain prediction, the diffusivity of the terminal fragments in
the presence of crowders for small $n$ shows sizeable deviations. 

The effective number of monomers involved in the looping dynamics $n_{\mathrm{
eff}}\propto D(1,\phi)/D_{\mathrm{eff}}(n,\phi)$ increases slightly with $n$ both for large and small
crowders, as shown in Fig.~\ref{fig8}. This figure illustrates that the number of chain monomers participating in looping slightly but systematically decreases with the MMC fraction $\phi$.
The functional dependence of $D_{\mathrm{eff}}$ is qualitatively similar to that of Rouse chains
at larger $n$, but with somewhat smaller $D_{\mathrm{eff}}$ values. 
For smaller
$n$, however, a plateau of $D_{\mathrm{eff}}$ is observed for all chain lengths in simulations.
For severe crowding we find that \emph{less} monomers are involved in looping,
compare the curves in Fig.~\ref{fig8}. This analysis rationalises the \textit{cooperativity}
between the polymer extremities and the vicinal crowding particles.
The $\phi$-dependent chain end diffusivity is our \emph{third main result}.
\footnote{Note that Eq. (\ref{eq-tl-fe-shape}) provides a satisfactory description of the looping times \cite{shin12looping}. Polymer looping is a prolonged barrier-crossing process in which the chain is close to equilibrium. For the reverse process of polymer unlooping, the disjoining of the end monomers takes place over a very short distance and spontaneous free-energy-downhill chain opening events occur (a process, which is inherently out of equilibrium). This is the main reason not to use the free energy-based Eq. (\ref{eq-tl-fe-shape}) to evaluate the times of chain unlooping. The latter consists of two terms, the time of disjoining the end monomers and their diffusion from a close distance to the separation $r_{\text{eq}}$. Depending on the attraction strength $\epsilon_s$, the relative contribution of the two terms to $T_{ul}$ varies. For large  $\epsilon_s$ values, for instance, the first contribution dominates so that the unlooping time exhibits the Arrhenius-like kinetics, see Eq. (\ref{eq-arrenius}) and Fig. \ref{fig9} below. }

\subsection{Effects of the binding affinity}

\begin{figure}
\includegraphics[width=8.8cm]{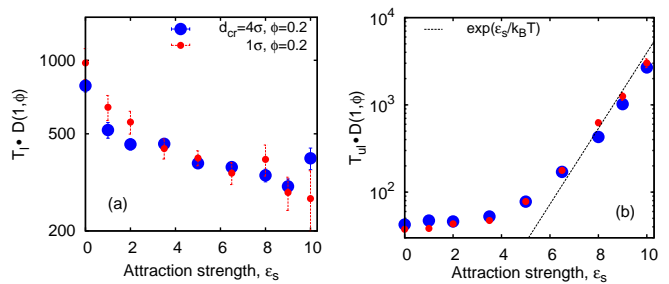}
\caption{Viscosity-renormalised looping (a) and unlooping times (b), namely $T_{l,ul}\to T_{l,ul}/\eta \sim T_{l,ul}D(1,\phi)$, at $\phi=0.2$
for varying $\epsilon_s$ and $n=32$. Note that with increase of $\epsilon_s$ the
number of looping events for the same trace-length of $r(t)$ decreases,
worsening the statistics.
\label{fig9}}
\end{figure}

For ssDNA hairpins, the enthalpy gain of base-pairing upon looping is partly
counter-balanced by the entropic penalty \cite{russians-hairpins-non-entropic,
RNA-folding-landscape-PNAS}. For
instance, for the 21-bp hairpin with CCCAA/GGGTT termini in Ref.~\cite{weiss13crowd}
the free energy of hairpin formation is $\sim 5 k_\text{B}T$ \cite{jost09}. This value is
used in our simulations for the end-to-end binding energy $\epsilon_s$, except for
Fig.~\ref{fig9} where we vary $\epsilon_s$ in the broad range $0 \leq \epsilon_s \leq 10 k_\text{B}T$. Larger $\epsilon_s$ values represent ssDNA hairpins with longer and thus more
adhesive complementary end sequences. We observe a moderate, monotonic decrease of
the looping time with $\epsilon_s$. Moreover, for all $\epsilon_s$ values longer
looping times are obtained for smaller crowders and faster looping is detected for
larger crowders (Figs. \ref{fig9} and \ref{figs10}). This implies that our claims regarding the
effects of crowding on the polymer looping dynamics are robust to changes of the
model parameters. 

Both looping and unlooping times can be rescaled by the effective
solution viscosity $\eta\sim1/D(1,\phi)$ to yield \emph{universal\/} dependencies for different
crowder sizes as demonstrated in Fig.~\ref{fig9}b. Fig.~\ref{figs10}
shows the unscaled looping and unlooping data, together with the results for in absencd of crowding, revealing the same trends for $T_{l,ul}$ with the crowder size as those presented for a fixed end-monomer affinity in Fig. \ref{fig5}b. The viscosity-based rescaling works particularly
well for the unlooping. 
As expected, for large end-to-end attraction
$\epsilon_s$ the unlooping time grows and exhibits Arrhenius-like kinetics, \begin{equation}T_l(\epsilon_s)
\sim\exp[\epsilon_s/(k_BT)],\label{eq-arrenius}\end{equation}see the dotted line in Fig.~\ref{fig9}b. These
findings regarding the binding strength are our \emph{fourth key result}. The exponential growth of the unlooping time with $\epsilon_s$ indicates the \textit{local }physical nature of the unlooping process, in contrast to the looping kinetics at varying attractive strength  $\epsilon_s$  which requires rather \textit{large-scale }polymer re-organisations.

\subsection{Cavity and Caging}
\label{sec-caging}

To quantify the already mentioned caging effects imposed by the crowders on the polymer coil, we explicitly compute the distribution of crowders around the polymer, as  illustrated in Fig. \ref{fig-crowder-distr-new}. It shows that crowding particles of size comparable to the chain monomers diffuse quite substantially inside the coil volume. In contrast, crowders, whose size is much larger than the polymer monomers, are essentially excluded/depleted from the volume occupied by the polymer, thus facilitating polymer compaction and looping. Here, the reader is also referred to the investigation of caging effects in colloidal glasses \cite{poon09pnas}.

We also evaluated the correlation characteristics of the number of contacts $m_{\text{cr-p}}(t)$ that the polymer chain establishes with the neighbouring crowders in the course of time, see Fig. \ref{fig-crowder-corr} for relatively large crowders. We define the normalised auto-correlation function of polymer-crowders contacts as \cite{shin14rings} \begin{eqnarray}\label{eq-ACF}
\nonumber
\text{ACF}(\Delta)=~~~~~~~~~~~~~~~~~~~~~~~~~~~~~~~~~~~~~~~~~~~~~~~~~~~\\
\frac{\left<m_{\text{cr-p}}(t+\Delta)m_{\text{cr-p}}(t)\right>-\left<m_{\text{cr-p}} (t+\Delta)\right>\left<m_{\text{cr-p}}(t)\right>}
{\left<m_{\text{cr-p}}(t)^{2}\right>-\left<m_{\text{cr-p}}(t)\right>^2}, \label{eq-acf}\end{eqnarray} where the angular brackets denote averaging along the $m_\text{cr-p}(t)$ trace. The critical distance between the centres of polymer monomers and neighbouring crowders in the algorithm is set to $R_c=\sigma/2+d_{\text{cr}},$ such that at most one crowder fits between a crowder and a polymer monomer in contact. We checked that the observed ACF($\Delta$) decay length is only weakly sensitive to the chosen critical contact distance $R_c$.

We observe that, after an initial fast decrease of the number of contacts established, the further decay of the correlation function becomes nearly exponential, ACF($\Delta)\sim\exp[-\Delta/T^\star]$, see Fig. \ref{fig-crowder-corr}. The corresponding decay length $T^\star$ increases for longer polymers, partly due to a larger number of overall contacts $m_{\text{cr-p}}$ established. The characteristic time scale $T^\star$  we obtain here is substantially \textit{shorter} than the polymer looping time $T_l$ at the same conditions, compare Fig. \ref{fig5} and Fig. \ref{fig-crowder-corr}.
\footnote{The reason is as follows. The decay time characterising ACF$(\Delta)$ is related to the dynamics of \textit{individual} monomer-crowder contacts. The interplay of these  local fluctuations defines the life-time of a \textit{cage }mediated by larger crowders for the entire polymer. The looping time is, on the other hand, a target-search problem for the encounter reaction of the two polymer ends at the same small region of space. Being impeded by a topological polymer structure chain looping takes place typically on much longer time scales than $T^\star$.}

\section{Results: Diffusion}
\label{sec-results-diffusion}

\subsection{Subdiffusion of polymer ends}

\begin{figure}
\begin{center}
\includegraphics[width=8.8cm]{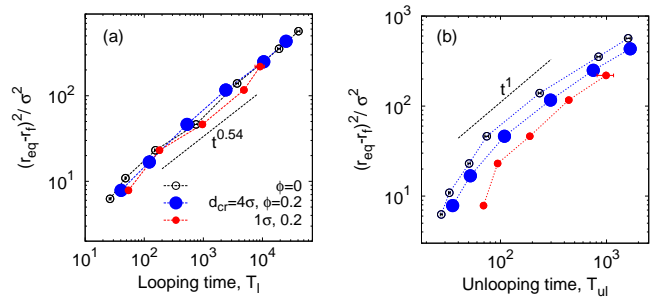}
\end{center}
\caption{Diffusion law of Eq.~\eqref{eq-EED} for looping and
unlooping times of the chain ends necessary to bridge the distance $\left<\delta
r(n)^2\right>$. The asymptote  of Eq.~\eqref{eq-tl-scaling} is the
dotted line in panel (a); a linear scaling in panel (b) as a guide for the
eye. Parameters are the same as in Fig. \ref{fig5}
and $\phi=0.2$. 
\label{figs8}}
\end{figure}

MMC may impede the folding dynamics of short polypeptides due to a higher solution
viscosity \cite{prot-fold-size} overwhelming the looping-facilitating caging effects. A
size-dependent diffusivity emerges \cite{prot-fold-size}: the diffusion of longer
chains is impeded more strongly. Fig.~\ref{figs8} based on our simulations shows a
similar effect for the mean squared looping distance versus the looping and
unlooping times. This quantifies the diffusion law for looping events, i.e., the
diffusive bridging of the distance \begin{equation}\delta r=r_{\text{eq}}-r_c \end{equation} from the equilibrium
distance $r_{\text{eq}}$ of the sticky ends to the looped state with end-to-end distance
$r_c$, and vice versa. As function of the chain length $n$, we checked that,
similar to $R_g$ in Fig.~\ref{fig3}b, for longer polymers the scaling law \begin{equation}\left<
\delta r(n)\right>^2\sim n^{2\nu}\end{equation} is fulfilled. From the mean times $T_l$ and
$T_{ul}$ we compute the scaling exponents $\alpha$ from the generalised diffusion
law \cite{RM-AD}
\begin{equation}
\left<\delta r(n)^2\right>=2D_{\alpha_l}\left<T_{l}(n)\right>^{\alpha_l}=2D_{
\alpha_{ul}}\left<T_{ul}(n)\right>^{\alpha_{ul}}.
\label{eq-EED}
\end{equation}
Here $D_{\alpha_i}$ is the generalised diffusion coefficient in units of $\mathrm{
cm}^2\mathrm{sec}^{-\alpha_i}$ and $\alpha_i$ the anomalous diffusion exponent for
looping and unlooping processes, respectively. This approach helps us to distinguish the
effects of the enhanced viscosity at higher $\phi$ from excluded-volume effects of
crowders. Fig.~\ref{figs8}a illustrates that at large $\phi$ the looping dynamics
is subdiffusive with $0.5\lesssim \alpha_l\lesssim0.6$. This is but the standard result for
polymer looping, as seen from combination of Eqs.~\eqref{eq-tl} and \eqref{eq-EED},
\begin{equation}
\left<\delta r(n)^2\right>\sim T_l(n)^{2\nu/(2\nu+1)}\sim T_l(n)^{0.54}.
\label{eq-tl-scaling}
\end{equation}
In contrast, for polymer unlooping no power-law scaling is found in the range of chain
lengths $n\sigma$ considered here, see Fig.~\ref{figs8}b. This fact is related to the absence of a power-law scaling in the $T_{ul}(n)$ dependence, see Fig. \ref{fig5}b.

These observations can be rationalised as follows. Once a thermal fluctuation
breaks the bond between the sticky ends, the separation $r$ of the polymer ends
drifts downhill in the free energy landscape $F(r)$ discussed above, quickly
assuming larger values. 
In contrast, the looping
time depends strongly on $n$: to loop, the polymer needs to overcome an entropic
penalty to get from $r_{\text{eq}}$ to the contact distance $r_c$, see Fig.~\ref{figs2}.
Thus, for looping it takes much longer to bridge the distance $\delta r
(n)$ and involves interactions with a larger number of surrounding crowders,
effecting the power law \eqref{eq-tl-scaling} with the small value $\alpha_l=
0.54$.

\begin{figure}
\begin{center}
\includegraphics[width=7cm]{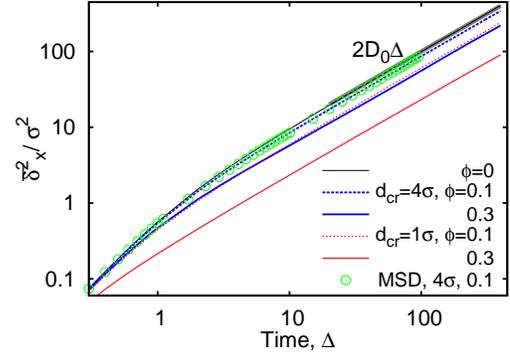}
\end{center}
\caption{MSD $\left<s^2(t)\right>$ (green symbols) and time averaged MSD
$\left<\overline{\delta^2_x(\Delta)}\right>$ along one spatial directions
computed for varying crowder size and MMC fractions $\phi$ (as indicated in
the plot). For each set of parameters, we average over $N=5$ time series for
$\left<\overline{\delta^2}\right>$ and over $N=10^3$ traces for the MSD. The
asymptote  MSD$_{x}(t)=2Dt$  is the thick solid line, where $D=D(1,0)$ is the
single monomer free space diffusivity.
\label{figs9}}
\end{figure}

\subsection{Diffusion of a tracer particle}

The size of the obstacles controls the facilitation or inhibition of polymer
looping in crowded environments. Additionally, we exploit how fast
the polymer ends join one another from the extended equilibrium state and
reveal the regime of anomalous diffusion for the looping times with the scaling
exponent of  $\approx$0.54, see Eq.~\eqref{eq-tl-scaling}. Here we briefly examine
whether this sub-diffusive behaviour of extended polymer
extremities is connected to
any subdiffusion of an \textit{isolated tracer particle }in the crowded solutions simulated.

We compute the mean square displacement (MSD) for the diffusion of a single monomer of the chain (tracer particle), $\left<s^2(t)\right>$, in crowded solutions with varying MMC fraction $\phi$ and crowder diameter $d_{\text{cr}}$. Namely, we use the anomalous diffusion law \cite{pt,RM-AD,RM-AD-2,rm-pccp2014} \begin{equation}
\left<s^2(t)\right>\sim t^\beta
\label{eq-s2-ad}
\end{equation}  to compute the local scaling exponent
\begin{equation}
\beta(t)=d[\log(\left<s^2(t)\right>)]/ d[\log(t)]
\label{eq-beta}
\end{equation}
along the ensemble averaged MSD trajectory. For the Brownian motion $\beta(t)\equiv 1$ at all times.

We find that the viscosity of solutions of smaller crowders grows with $\phi$
faster than for larger obstacles, see Fig. \ref{figs7}. In this figure, the
diffusivity has been extracted from the time averaged MSD along $x-$direction,
\begin{equation} \overline{\delta^2_{x,i}(\Delta)}= \frac{1}{t-\Delta}
\int_0^{t-\Delta}[x_{i}(t'+\Delta)-x_{i}(t')]^2dt', \label{eq-tamsd}\end{equation} in
the lag time interval of $40<\Delta<400$, i.e., in the region where the  linear scaling of the time average MSD is clearly established. This fast increase of the
tracer's viscosity is consistent with the experimental measurements in crowded
dextran solutions, see Fig. 5b in Ref. \cite{dextran-expon-diffusion}. In the
latter, the tracer exhibits an exponential growth of micro-viscosity with
 concentration of polymeric crowders, valid for a wide range of relative tracer-crowder
dimensions. The growth of viscosity with MMC fraction $\phi$ is also in
accord with theoretical predictions \cite{minton12viscosity}.

The MSD and ensemble averaged time averaged MSD $\left<\overline{\delta^2_x(\Delta)}\right>=N^{-1}\sum_{i=1}^N \overline{\delta^2_{x,i}(\Delta)}$ traces are identical in the long-time
limit, see Fig. \ref{figs9}, with the long-time exponent $\beta$ being close to unity
(Brownian motion). This indicates the \textit{ergodic} tracer diffusion in the crowded
solutions implemented in our simulations yields subdiffusive motion of the chain
ends. In Fig.~\ref{figs9} we also show the ensemble and time averaged MSDs of a tracer particle with unit diameter in the crowded
solutions. The diffusion exponent is nearly unity and no disparity of ensemble
and time averaged displacements is detected, i.e., the motion is ergodic \cite{pt,rm-pccp2014}.

\section{Discussion}
\label{sec-discussion}

MMC non-specifically favours more compact conformations of proteins and speeds up
their folding kinetics \cite{minton08}, as well as stabilises the proteins against
thermal denaturation \cite{minton82}. MMC may also reduce the occurrence of
mis-folded states via reduction of the conformational space \cite{thirum05}. The
degree of crowding in living cells is heterogeneous and the crowders are polydisperse in size \cite{surya-polydisperse}, giving
rise to a micro-compartmentalisation of the cellular cytoplasm \cite{microcomp-1,
microcomp-2,microcomp-3}. These effects pose the questions whether other
fundamental elements of gene expression in biological cells are equally affected
by MMC. 

Specifically, recent gene-regulation experiments \cite{leduc13} have shown that bigger
dextran molecules increase the rates of gene expression by RNA Polymerase to
a higher-fold as compared to smaller ones \cite{leduc13}. Bigger dextran
molecules both reduce the diffusivity of RNA Polymerase and enhance the
number of binding events to the promoters  (enhancing the association and
reducing the dissociation rates). In solution  of small crowders the impact
of $\phi$ on gene expression rates is non-monotonic (due to a compensation
of moderate effects of MMC on Polymerase diffusivity and its association
rate to the DNA sites). In contrast, in solution of bigger crowders the
expression rate grows monotonically and strongly with the $\phi$ fraction \cite{leduc13}.

Here we show that indeed the looping kinetics of polymers such as DNA is
highly sensitive to the volume fraction and size of crowders in a non-trivial
way, and a quantitative knowledge of this effect is necessary for the understanding
of the molecular biological function of DNA
based on looping. From extensive Langevin dynamics simulations we  demonstrated
that polymer looping is facilitated in the presence of large crowders, mainly due to depletion-based chain compaction.
In contrast, for small crowders the dominant effect is the larger effective
viscosity impeding the looping dynamics. The exact tradeoff between the two effects
critically depends on the system parameters.

Our results are applicable to generic DNA looping and RNA folding dynamics
in crowded systems \cite{DNA-folding-MMC}, particularly, the formation
of ssDNA hairpins with in vitro crowders \cite{weiss13crowd}.
Here, our predictions for crowder size and binding affinity effects
can be tested directly in experiments. We already have showed that some predictions of our model indeed
capture the experimental behaviour \cite{weiss13crowd}.
As targets for future studies, crowders of
particular surface properties, non-inert poly-disperse and aspherical crowders
will be studied \cite{looping-future}. 

In addition, the simulations of
semi-flexible instead of flexible polymers in the presence of both MMC and external
spherical confinement are expected to reveal a number
of novel features. For instance, in contrast to free-space flexible chains, the
presence of spacial restrictions and finite bending energy penalty upon polymer looping yields a
quasi-periodic but \textit{highly erratic }dependence with the chain length $n \sigma$.
Strong anti-correlation of the looping time and looping probability versus the
polymerisation degree, pertinent for flexible chains, as those presented in
Figs.~\ref{fig3}a and \ref{fig5}a, become more profound for the dynamics of
cavity-confined semi-flexible polymers, see Ref. \cite{looping-future}. We hope that our current investigation triggers new theoretical and experimental developments of static and dynamical properties of polymers in the crowded realm omnipresent in the interior of living cells.
\footnote{After submitting the current manuscript, we became aware of the recent studies of the crowder size \cite{pincus-prl,kor-new-sm}. A stabilisation of intrinsically-disordered proteins and stabilisation of coil-to-globule transitions by crowding was discussed in Ref. \cite{pincus-prl}, based on computer simulations of an MMC-induced compaction of polymers. It was shown e.g. that smaller crowders exerting a higher osmotic pressure onto the polymer compact it to a larger extent, as compared to the bigger ones. Contrary to our observations, particularly small crowders are excluded from the space occupied by the self-avoiding polymer. Similar to our results, Ref. \cite{pincus-prl} indicated that the size of the polymer coil reduces monotonically with $\phi$. A slight non-monotonic $R_g(\phi)$ dependence obtained for the same system \cite{} based on a phenomenological depletion potentials \cite{loop-crowd-simul-1} is thus rendered to be an artifact \cite{pincus-prl}. The effects of MMC in our system are weaker than in Ref. \cite{pincus-prl} (we have the Flory-like scaling of polymer dimensions and no coil-to-globule transitions occur). The difference may be due to a smaller size of crowders in Ref. \cite{pincus-prl}, as compared to the polymer monomers. Similarly to our results presented in Fig. \ref{figs1}b, in Ref. \cite{slater-last} smaller crowders were shown to be  more efficient in compacting the polymer chain. Lastly, in Ref. \cite{kor-new-sm} the effects of the crowder size was investigated regarding the strength of depletion interactions between the \textit{two} polymers. The strength of effective polymer-polymer attraction was shown to be reduced as the crowder size decreases (at a constant $\phi$ fraction).}

\begin{acknowledgments}
We thank D. Jost for discussions. We acknowledge funding from the Academy of
Finland (FiDiPro scheme to RM), the Deutsche Forschungsgemeinschaft (DFG Grant
CH 707/5-1 to AGC), and the Federal Ministry of Education and Research (BMBF
Project to JS). We are particularly grateful to an anonymous
referee for the insightful comments which improved our understanding of the 
subject.

\end{acknowledgments}

\begin{appendix}

\section{}
\numberwithin{figure}{section}
\setcounter{figure}{0}

In this Appendix we present the supplementary figures explaining the details of our main-text results.

\begin{figure}
\begin{center}
\includegraphics[width=6.5cm]{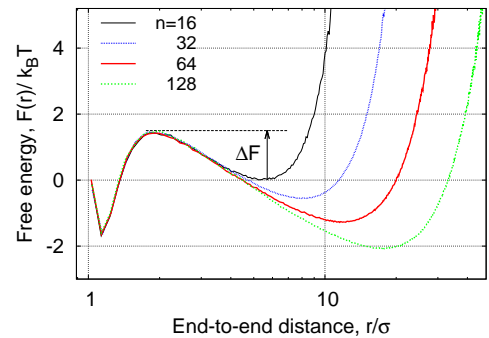}
\end{center}
\caption{The free energy landscape $F(r)$ for polymer looping for varying
chain length at  $\phi=0.2$ and  $d_{\text{cr}}=4\sigma$. The energy minima at
small end-to-end distances are aligned in the plot in order to assess the barriers
heights for looping, $\Delta F(n)$.
\label{figs2}}
\end{figure}

\begin{figure}
\begin{center}
\includegraphics[width=6cm]{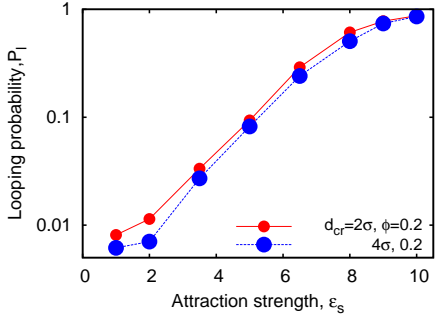}
\end{center}
\caption{Looping probability versus terminal monomer stickiness, computed for $n=32$
chains at different crowder size at $\phi=0.2$.
\label{figs4}}
\end{figure}

\begin{figure}
\begin{center}
\includegraphics[width=9cm]{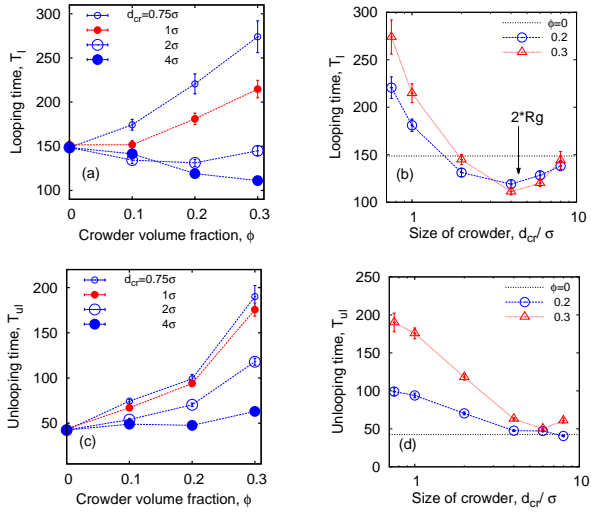}
\end{center}
\caption{Looping and unlooping times versus  $\phi$, computed for a varying crowder
size $d_{\text{cr}}$, for short chains with $n=16$ monomers, and $\epsilon_s=5k_\text{B}T$.
\label{figs3}}
\end{figure}

\begin{figure}
\begin{center}
\includegraphics[width=8cm]{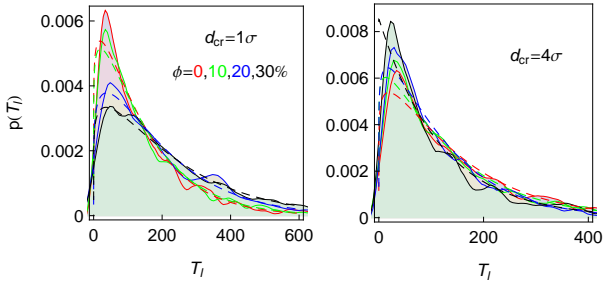}
\end{center}
\caption{The PDFs of looping times for $n=16,~\epsilon_s=5k_\text{B}T$ and crowder sizes and fractions as
indicated. Smoothed histograms are the simulations data and the dashed curves of the respective colour are the fits by
Eq.~\eqref{eq-weibull}. The slight bumpiness of the histograms is due to the limited statistics of the generated looping events.
\label{figs5}}
\end{figure}

\begin{figure}
\begin{center}
\includegraphics[width=6cm]{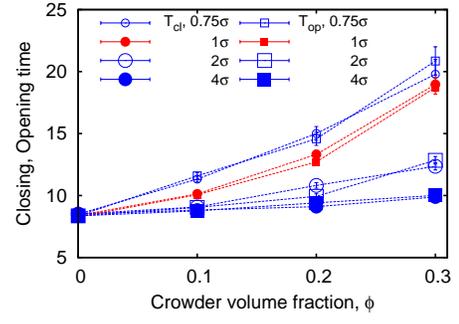}
\end{center}
\caption{Opening and closing times $T_{op,cl}$ versus MMC fraction
$\phi$ for varying crowder size. Parameters are the same as in
Fig. \ref{figs3}.
\label{figs6}}
\end{figure}

\begin{figure}
\begin{center}
\includegraphics[width=6cm]{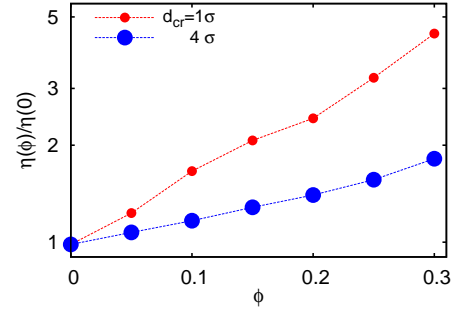}
\end{center}
\caption{Effective solution viscosity $\eta(\phi)=k_\text{B}T/(3\pi\sigma D(1,\phi))$
for a tracer of diameter $1\sigma$ in solutions with varying crowder
diameter $d_{\text{cr}}$, as extracted from the analysis of time averaged MSD traces. 
\label{figs7}}
\end{figure}

\begin{figure}
\includegraphics[width=6.5cm]{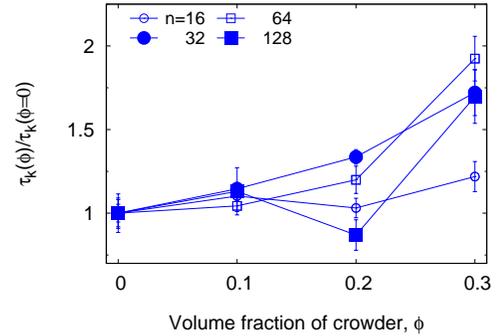}
\caption{The normalised $\tau_K(\phi)$ for varying chain length $n$, plotted for $\epsilon_s=5k_\text{B}T$ end-monomer adhesion strength and relatively big crowders $d_{\text{cr}}=4\sigma$. For small crowders the effect $\phi$ on the enhancement of $\tau_K$ is stronger and more systematic (not shown).
\label{figs7a}}
 \end{figure}

\begin{figure}
\begin{center}
\includegraphics[width=8cm]{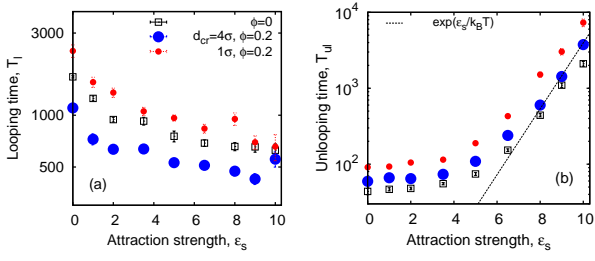}
\end{center}
\caption{The same data as in Fig. \ref{fig9} but without normalisation by
the effective viscosity of the solution, $\eta(\phi)\sim 1/D(1,\phi)$. The data-set for the uncrowded solution
is also included (open symbols).
\label{figs10}}
\end{figure}

\begin{figure}
\includegraphics[width=7cm]{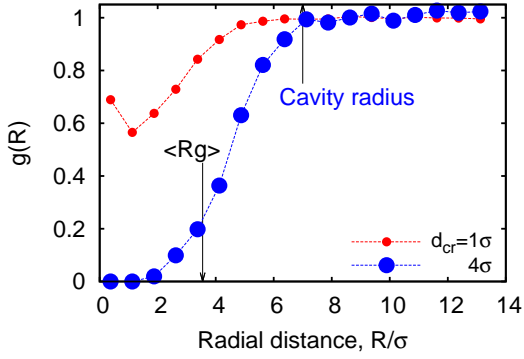}
\caption{The radial distribution function of relatively small (red dots) and large (blue dots) crowders around a polymer coil with $n=32$ monomers at MMC fraction of $\phi=0.1$.}
\label{fig-crowder-distr-new}
\end{figure}

\begin{figure}
\includegraphics[width=7cm]{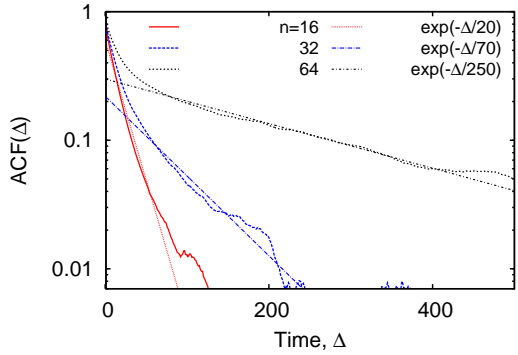}
\caption{The auto-correlation function of the polymer-crowders contact number  (\ref{eq-acf}), computed for polymers of varying length, at $\phi=0.2$ and $d_{\text{cr}}=4\sigma$. The corresponding exponential asymptotes are shown as the dotted lines.}
\label{fig-crowder-corr}
\end{figure}

\end{appendix}

\clearpage

\bibliographystyle{apsrev4-1}

\end{document}